%% file: main.tex
\documentclass[acmsmall]{acmart}

\usepackage{xspace}
\usepackage{listings}
\usepackage{multirow}
\usepackage{subcaption}
\usepackage{arydshln}

\newcommand{\eg}{{\it e.g.,\ }}
\newcommand{\etal}{{\it et al.\ }}

\newcommand{\ie}{{\it i.e.,\ }}

\newcommand{\aka}{{a.k.a.}\xspace}
\newcommand{\tool}{\textit{TranSlider}}

\newlength{\tboxsize}
\newcommand{\tbox}[1]{%
  \fbox{%
    \parbox[c][\tboxsize][c]{\tboxsize}{%
      \centering      
      #1              
    }%
  }%
}

\newcommand{\BoxedAppendix}[1]{%
  \begin{center}
    \setlength{\fboxsep}{9pt}
    \fcolorbox{black}{gray!10}{%
      \begin{minipage}{0.9\textwidth}
        \ttfamily\small
        #1
      \end{minipage}%
    }
  \end{center}
}

\AtBeginDocument{%
  }

\setcopyright{cc}
\setcctype{by-nd}
\acmJournal{PACMHCI}
\acmYear{2025} \acmVolume{9} \acmNumber{7} \acmArticle{CSCW479} \acmMonth{11} \acmPrice{}\acmDOI{10.1145/3757660}

\begin{document}

\title{Steering AI-Driven Personalization of Scientific Text for General Audiences}


\author{Taewook Kim}
\authornote{Work done during an internship at Accenture Labs, San Francisco}
\affiliation{%
  \institution{Northwestern University}
  \city{Evanston}
  \state{IL}
  \country{USA}}
\email{taewook@u.northwestern.edu}

\author{Dhruv Agarwal}
\authornotemark[1]
\affiliation{%
  \institution{Cornell University}
  \city{Ithaca}
  \state{NY}
  \country{USA}}
\email{da399@cornell.edu}

\author{Jordan Ackerman}
\authornote{Work done during their tenure at Accenture Labs}
\affiliation{%
  \institution{Center for Advanced AI, Accenture}
  \city{San Francisco}
  \state{CA}
  \country{USA}}
\email{jordan.ackerman@accenture.com}

\author{Manaswi Saha}
\affiliation{%
  \institution{Accenture Labs}
  \city{San Francisco}
  \state{CA}
  \country{USA}}
\email{manaswi.saha@accenture.com}

\renewcommand{\shortauthors}{Taewook Kim, Dhruv Agarwal, Jordan Ackerman, and Manaswi Saha}

\begin{abstract}
Digital media platforms (\eg science blogs) offer opportunities to communicate scientific content to general audiences at scale. However, these audiences vary in their scientific expertise, literacy levels, and personal backgrounds, making effective science communication challenging. To address this challenge, we designed \textit{TranSlider}, an AI-powered tool that generates personalized translations of scientific text based on individual user profiles (\eg hobbies, location, and education). Our tool features an interactive slider that allows users to steer the degree of personalization from 0 (weakly relatable) to 100 (strongly relatable), leveraging LLMs to generate the translations with chosen degrees. Through an exploratory study with 15 participants, we investigated both the utility of these AI-personalized translations and how interactive reading features influenced users' understanding and reading experiences. We found that participants who preferred higher degrees of personalization appreciated the relatable and contextual translations, while those who preferred lower degrees valued concise translations with subtle contextualization. Furthermore, participants reported the compounding effect of multiple translations on their understanding of scientific content. Drawing on these findings, we discuss several implications for facilitating science communication and designing steerable interfaces to support human-AI alignment.

\end{abstract}


\received{October 2024}
\received[revised]{April 2025}
\received[accepted]{August 2025}

\maketitle

\input{1-introduction}
\input{2-relatedwork}
\input{3-system}
\input{4-study}
\input{5-results}
\input{6-discussion}
\input{7-conclusion}


\begin{acks}
We thank all reviewers for their constructive feedback. We also thank Accenture Labs' leadership for supporting this research and Labs' Digital Experiences team members for continued guidance. Furthermore, special thanks to members from the broader research community, including Elizabeth Churchill, Jung Wook Park, John Chung, Duri Long, and Tal August for their insightful discussion. We also greatly appreciate Matt Kay for his thoughtful comments and suggestions on the manuscript and Jessica Hullman for introducing extensive related work in this domain.
\end{acks}

\bibliographystyle{ACM-Reference-Format}
\bibliography{reference}

\appendix

\section{Original Science Articles}

\subsection{Health~\cite{kong2024glycolytic}}

\BoxedAppendix{%
  \textbf{Title:} A glycolytic metabolite bypasses ``two-hit'' tumor suppression by BRCA2\\[4pt]
  \textbf{Abstract:} Knudson's ``two-hit'' paradigm posits that carcinogenesis requires inactivation of both copies of an autosomal tumor suppressor gene. Here, we report that the glycolytic metabolite methylglyoxal (MGO) transiently bypasses Knudson's paradigm by inactivating the breast cancer suppressor protein BRCA2 to elicit a cancer-associated, mutational single-base substitution (SBS) signature in nonmalignant mammary cells or patient-derived organoids. Germline monoallelic BRCA2 mutations predispose to these changes. An analogous SBS signature, again without biallelic BRCA2 inactivation, accompanies MGO accumulation and DNA damage in Kras-driven, Brca2-mutant murine pancreatic cancers and human breast cancers. MGO triggers BRCA2 proteolysis, temporarily disabling BRCA2's tumor suppressive functions in DNA repair and replication, causing functional haploinsufficiency. Intermittent MGO exposure incites episodic SBS mutations without permanent BRCA2 inactivation. Thus, a metabolic mechanism wherein MGO-induced BRCA2 haploinsufficiency transiently bypasses Knudson's two-hit requirement could link glycolysis activation by oncogenes, metabolic disorders, or dietary challenges to mutational signatures implicated in cancer evolution.
}

\subsection{Environment~\cite{Guelfo2024}}

\BoxedAppendix{%
  \textbf{Title:} Lithium-ion battery components are at the nexus of sustainable energy and environmental release of per- and polyfluoroalkyl substances\\[4pt]
  \textbf{Abstract:} Lithium-ion batteries (LiBs) are used globally as a key component of clean and sustainable energy infrastructure, and emerging LiB technologies have incorporated a class of per- and polyfluoroalkyl substances (PFAS) known as bis-perfluoroalkyl sulfonimides (bis-FASIs). PFAS are recognized internationally as recalcitrant contaminants, a subset of which are known to be mobile and toxic, but little is known about environmental impacts of bis-FASIs released during LiB manufacture, use, and disposal. Here we demonstrate that environmental concentrations proximal to manufacturers, ecotoxicity, and treatability of bis-FASIs are comparable to PFAS such as perfluorooctanoic acid that are now prohibited and highly regulated worldwide, and we confirm the clean energy sector as an unrecognized and potentially growing source of international PFAS release. Results underscore that environmental impacts of clean energy infrastructure merit scrutiny to ensure that reduced CO2 emissions are not achieved at the expense of increasing global releases of persistent organic pollutants.
}

\end{document}

%% file: 1-introduction.tex
\section{Introduction}

Science communication refers to the process of ``communicating complex scientific information with \textit{general audiences}\footnote{We refer to ``general audiences'' as individuals who are non-scientists, coming from diverse educational backgrounds and domains such as arts, business, education, healthcare, and product design.} to improve public awareness, interest, and understanding of science''~\cite{burns2003science}. Both academic and industry professionals strive to make their ideas more accessible. As a result, science communication platforms, such as science blogs~\cite{sciencedaily} and science magazines~\cite{science}, have been translating scientific information into more accessible formats for decades. However, a single version of a text often falls short of meeting the diverse needs of a general audience~\cite{august2024know}. For instance, a journalist with an interest in quantum physics would have vastly different context and comprehension needs than a college student majoring in physics.

Scientific research articles often employ domain-specific language, including jargon and complex sentence structures, which can pose significant barriers to comprehension for general audiences. Language models offer the capability to personalize content to various user contexts, transforming both the style (\eg~casual to formal~\cite{das2023balancing} or romantic~\cite{kim2019love}) and content (\eg~simplifying complex ideas~\cite{august2024know}). This adaptive personalization creates opportunities for enhancing audience engagement and understanding through more effective content dissemination~\cite{kim2024authors}.
Notably, large language models (LLMs) can achieve this personalization at scale (\ie~AI-scalable personalization), generating multiple tailored variations of content for diverse audiences~\cite{august2024know, riedl2010scalable, kim2024authors}. This scalability positions LLMs as powerful tools in advancing inclusive and effective science communication~\cite{razack2021AI}.

Among the many strategies in science communication from prior HCI/CSCW literature, analogies have proven effective in translating technical content into more accessible forms~\cite{august2020writing}. For instance, the structure of the solar system is often used as an analogy to explain the structure of an atom. 
While such analogies can broaden understanding of scientific information, they are limited in number, can be generic, and may not resonate equally well with all audiences as individual comprehension is shaped by societal, cultural, educational, and personal backgrounds. For example, the space analogy assumes familiarity with the solar system~\cite{kim2024authors}. LLMs have demonstrated the ability to generate personalized analogies that can help people understand complex concepts and ideas within their own context~\cite{ding2023fluid}. Building on this capability, we leverage LLMs to generate multiple personalized analogies tailored to individual readers' contexts. 

We designed and implemented \tool{} (\textit{Trans}late through \textit{Slider}), an interactive reading interface that enables users to steer the degree of personalization in scientific text through an adjustable slider. \tool{} allows the user to specify a degree of personalization from 0 to 100 and utilizes their background information (\eg education, hobbies, location) to present relevant analogies. The slider enables intuitive exploration of various personalization degrees, allowing users to quickly review multiple translations. We employed \tool{} as a research probe to ask the following research questions:
\begin{itemize}
    \item[\textbf{RQ1:}] What is the utility of AI-driven personalized translations of scientific text?
    \item[\textbf{RQ2:}] What is the impact of interactive reading features on user experience? Specifically:
    \begin{itemize}
        \item[\textbf{2.1:}] How does exploring multiple translations influence readers' comprehension and engagement with scientific text?
        \item[\textbf{2.2:}] How does the slider interaction to steer the degree of personalization influence readers' comprehension and engagement with scientific text?
    \end{itemize}
\end{itemize}
To answer these questions, we conducted a user study with 15 non-expert participants who used the tool to understand two scientific texts by exploring multiple personalized translations. We conducted post-session semi-structured interviews to elicit feedback on their experience. We used thematic analysis on their qualitative responses and conducted descriptive quantitative analysis on tool usage logs to understand their exploration behavior (\eg number of generated translations. range of explored personalization degrees).

Participants found the analogies useful in understanding the content. Some favored general analogies (\eg a participant liked a construction analogy describing the body's cells as building sites---low personalization), while others preferred detailed, personalized analogies (\eg a participant appreciated a baking analogy to explain harmful emissions from lithium-ion batteries---high personalization). Participants noted that reading multiple translations with varied analogies allowed them to piece together a fuller understanding of the scientific text, correcting misunderstandings along the way.
Participants found the tool beneficial for learning about unfamiliar topics but were cautious about the reliability of AI-generated translations. Based on these findings, we discuss implications for science communication and the need for interactive techniques to steer models towards human pluralistic preferences, rather than solely relying on machine learning approaches.

Overall, our contributions to the CSCW community include,
\begin{enumerate}
    \item A novel slider-based LLM interaction to enhance understanding of scientific content through the exploration of analogy-driven personalized translations
    \item An investigation to examine the utility, benefits, and limitations of AI-driven personalized translations in science communication
    \item Implications for science communication and more broadly for cross-disciplinary communication, and designing interfaces for human-AI alignment.
\end{enumerate}

%% file: 2-relatedwork.tex
\section{Related Work}
We review prior work covering science communication on digital media platforms, personalization in science communication, and existing interactive reading interfaces for scientific articles.

\subsection{Science Communication on Digital Media Platforms}
\label{rw:scicomm}

Science communication takes many forms in modern media for communicating complex scientific information to general audiences for improving public awareness, interest, and understanding of science~\cite{burns2003science}. Traditional channels include museums~\cite{bell2008engaging}, exhibitions~\cite{long2021codesign}, and TV series~\cite{sagan1980cosmos}, while web-based digital platforms have emerged as a powerful new stream. For example, online videos have become an influential source for science learning, especially during the COVID-19 pandemic, when many people turned to digital resources for scientific information~\cite{Breslyn2022}. Social media platforms such as Reddit (\eg r/science), Twitter (now X), Bluesky, and Mastodon offer interactive spaces for scientists to share research and engage with the public~\cite{gero2021what, martin2020using, jones2019rscience}. Finally, science blogs have been valuable channels for explaining complex scientific content to general audiences~\cite{paige2018science}. For example, more than half of college students actively use blogs as a learning resource~\cite{head2017why}. A key advantage of science communication via digital media, \aka~online science communication~\cite{williams2022hci}, is its ability to reach a wider audience (\eg~\cite{nguyen2024simulating, jones2019rscience}). 

Prior HCI/CSCW literature has uncovered several key challenges scientists face in online science communication: engagement, translation, and dissemination~\cite{williams2022hci}. First, scientists rarely participate in digital media platforms~\cite{Zhu_Purdam_2017}. This limited engagement is due to the conventional academic nature that has put scientists as neutral observers rather than active communicators~\cite{Martinez-Conde2077, hans2013gap}, and practical constraints such as limited time and lack of communication training~\cite{besley2015besley}. While recent research shows a gradual shift towards more active public engagement by scientists~\cite{bruggemann2020post}, the extent and impact of this change remain to be fully understood. Secondly, even for scientists who engage with digital communication, translating complex scientific concepts into accessible language is challenging~\cite{ronald2017context}. These challenges may be compounded by the variety of readers with different backgrounds, levels of scientific literacy, and preferences~\cite{mike2017how, leona2014inequalities, ronald2017context}.
Creating multiple versions tailored to different expertise levels is infeasible for individual scientists~\cite{bucchi2014routledge, baruch2013sciences}. Lastly, translation challenges can entail broader concerns such as misinformation on social media platforms~\cite{cook2018benefits}. The dissemination of inaccurate interpretations or misrepresentations of their work could have unwanted consequences for public understanding and their professional reputation~\cite{wu2019design, maria2018responsibility}.

Given these challenges, this paper focuses on addressing the translation barrier in personalized science communication on digital media platforms. Specifically, we explore how to facilitate creating multiple translations of scientific content to effectively accommodate general audiences with varying levels of scientific background at scale.

\subsection{Personalization in Science Communication: Tools and Approaches}
\label{rw:personalization}

In science communication, personalization is an approach that translates scientific content into contexts familiar to each individual, making it more interesting, accessible, and engaging for non-scientists~\cite{anand2007intro}. For example, when explaining biological concepts to individuals without a background in biology, saying ``try not to drink soda and eat candy'' may be more effective than ``reduce glucose consumption'', as the former uses more familiar contexts while conveying the same information to the intended audience~\cite{li2024content}. Given its potential effectiveness for science communication, researchers have explored various strategies for personalizing information across contexts.

Prior approaches in HCI and CSCW have personalized scientific content using structured templates, often yielding limited benefits to the reader.
For instance, Persalog personalizes news articles by substituting specific segments---such as city, state, or country names---based on geographical relevance~\cite{adar2017persalog}. Other studies have employed template-based rephrasings. For instance, Hullman~\etal and Kim~\etal generate alternative expressions of measurements following a template: \texttt{``$\{$number$\}$ lb is about (X.X) times the weight of $\{$familiar object$\}$''}~\cite{hullman2018improving} or \texttt{``The distance between locations is (number) times your distance to a $\{$familiar landmark$\}$''}~\cite{kim2016generating}). These methods rely on structured databases, such as geographical information~\cite{adar2017persalog}, landmark references~\cite{kim2016generating}, and patients corpora~\cite{dimarco2006authoring}, to compute contextual relevance. While some work has attempted more flexible language generation (\eg~personalized medical brochures~\cite{dimarco2006authoring}), these efforts result in unnatural and ill-formed sentences, limiting their effectiveness.

These approaches predate the emergence of LLMs that could address these limitations: reliance on structured templates and databases. LLMs can transform text at scale without any databases, making it feasible to generate multiple translations on behalf of scientists. Recent HCI studies have explored the feasibility of LLMs in simplifying complex ideas (summarization)~\cite{august2023paper, gu2024an, kim2015descipher}, retargeting concepts into another domain (contextualization)~\cite{ding2023fluid, bao2025words, liu2025exploring}, and explaining content through user-familiar contexts (personalization)~\cite{august2024know, kim2024authors}. Albeit in other domains, these studies show that LLMs can reduce the reliance on databases to compute contextual relevance and generate natural-sounding language expressions tailored to various user contexts. Building on these works, our study explores integrating contextual personalization with LLMs' capability to generate analogies~\cite{ding2023fluid}. We aim to design an LLM-powered tool that leverages readers' contextual information to generate personalized analogies for scientific texts.

\subsection{AI-Powered Reading Interfaces for Scientific Articles}

Recent research has proposed AI-powered interactive reading interfaces for science articles. These systems summarize sections of the paper and help navigate within the article to efficiently find relevant information~\cite{august2023paper, fok2023scim}. For example, ScholarPhi assists users in quickly looking up definitions of nonce words and symbols defined elsewhere in the paper~\cite{head2021augmenting}. Researchers have also explored more free-form question-answering systems to help users identify relevant sections of a paper based on their queries~\cite{zhao-lee-2020-talk}. However, these primarily cater to users already motivated to engage with scientific literature, such as healthcare stakeholders (\eg patients~\cite{kambham2024explain}), students, and scientists.

While general-purpose tools like ChatGPT or AI-enabled PDF readers (\eg~Adobe Acrobat with built-in chatbots) can assist with reading scientific texts, they are not designed for scientific content, and require users to know what to ask. In contrast, our tool focuses on everyday readers with a casual interest in science---individuals who may lack the time, background, or motivation to actively engage with science articles. We position our tool as an AI-powered intermediary~\cite{kim2024authors} that delivers accessible, engaging science communication through analogy-driven, personalized translations. Rather than expecting readers to extract meaning themselves, our tool meets them where they are by adapting content to their context.

%% file: 3-system.tex
\section{Interaction Paradigms}

\subsection{Existing Design: Model-Initiative}
\label{model-init}
Existing personalization systems predominantly employ a model-initiative design, where AI models automatically collect and process user data to deliver personalized content without explicit user intervention~\cite{Corrigan01072014}. While efficient, this interaction design presents two primary human-AI alignment issues, namely personality alignment and user privacy.

\subsubsection{Personality Alignment Issues}
\label{alignment}
Model-initiative approach faces the \textit{personality alignment} issue~\cite{zhu2025personality},~\ie the personalized outcomes are not always well aligned with users' actual interests or needs~\cite{zollo2025personalllm, dudy-etal-2021-refocusing}. For example, consider an AI system that automatically personalizes results based on a user's current location (\eg Tokyo, Japan) when the user is actually interested in contextualized results of a different location (\eg Bergen, Norway). In this case, the AI personalization is misaligned with the user's actual interests. This alignment problem is compounded by the contextual nature of personalization preferences. Individuals may desire different levels of personalization depending on the context and motivation of their reading~\cite{august2024know, Grunig1979, leerink2024, dudy-etal-2021-refocusing}. For example, someone might prefer highly personalized content when reading medical articles relevant to their health conditions~\cite{Lim2011}, but relatively less or lower personalization when casually exploring scientific topics of general interest. To accommodate varying contextual needs for each individual at scale, the model-initiative approach would naturally require invasive data collection and inference about users' contexts.

\subsubsection{Privacy Issues}
\label{privacy}
Model-initiative design assumes that AI models have access to users' private data~\cite{Corrigan01072014}. However, from a user's perspective, it is unclear how much information is gathered, what types of information is gathered, and when~\cite{dudy-etal-2021-refocusing}. Furthermore, users typically have minimal control over their personal data~\cite{lenhart2023you}. These privacy concerns become more problematic in the context of science communication, which aims to reach diverse, mass audiences~\cite{burns2003science}. The model-initiative paradigm would require collecting and processing private data from the entire population of potential readers to provide personalized translations of scientific texts. This approach not only raises ethical concerns but also creates significant barriers to adoption, as many people would be reluctant to share their personal information.

\subsection{Our Design: Controlled User-Initiative}

We explore an alternative approach: user-initiative design. Existing LLM interfaces, such as ChatGPT~\cite{chatgpt} and Claude~\cite{claude3.5}, provide prompt interfaces where users can specify all personalization parameters and their intentions. While a prompt-based approach maximizes user control, it presents challenges specifically for science communication: allowing unlimited user modification risks compromising the scientific accuracy crucial for effective information dissemination. High degrees of freedom could lead to the distortion of key ideas or inadvertently promote misinformation, even when users have no intention to mislead~\cite{zhang2025unwanted, kim2024authors}.

We propose a controlled user-initiative design where readers can explicitly adjust specific aspects for personalization. To minimize the risk of misinformation or hallucination, rather than allowing free-form prompt interactions that could introduce unexpected results, we constrained user inputs to two structured elements: i) steerable \textbf{degrees} of personalization using sliders, addressing the alignment issue, and ii) editable structured \textbf{type} of contextual information via user profiles with predefined fields to alleviate the privacy issue. This design philosophy aims to preserve privacy by minimizing detailed personal data collection and granting users control over input, while enabling contextually relevant personalization that aligns with their needs and preferences.

\begin{figure}
  \includegraphics[width=\textwidth]{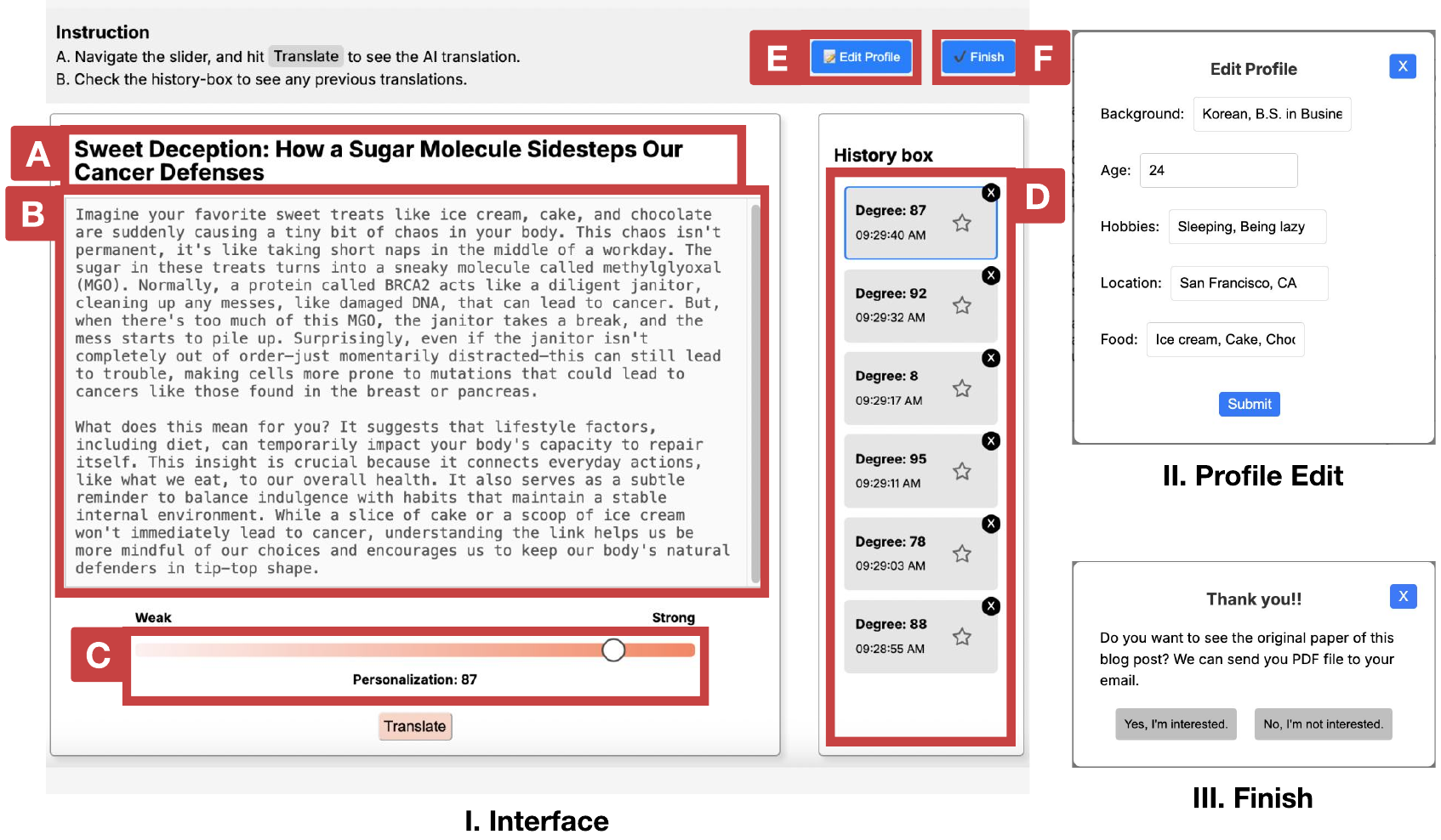}
  \caption{The interface design of \tool{}. \tbox{A} The AI-personalized title of the science blog post appears here. \tbox{B} AI-personalized translations of science articles will appear here. \tbox{C} Users can navigate the slider to adjust their preferred degree of AI personalization for a science article intro. \tbox{D} For each personalized translation, a history item is stacked in the history box. Users can revisit any previous history items to see the translations. \tbox{E} Users can click the edit profile button to see the Profile Edit window [II] to adjust their profile information. \tbox{F} The Finish button triggers a pop-up [III] to collect the user's interest in reading the original paper. We use the response as a proxy measure for their interest in the content after reading the personalized translations.}
  \label{fig:slider}
\end{figure}

\section{TranSlider: Design and Implementation}
\label{system}
With this design philosophy in mind, we designed and implemented \tool{} (\textit{Trans}late through \textit{Slider}), an LLM-powered reading interface to help general audiences better understand scientific texts by generating personalized translations through a slider interaction and user profiles.

\subsection{Interface Design}

We designed our tool with three main components: the translation panel, the history box, and the edit profile button (\autoref{fig:slider}). The header provides instructions on using the interface; these instructions were brief as participants received a detailed tutorial at the start of the study. 

\subsubsection{Translation Panel}
The translation panel has the main personalization features (\autoref{fig:slider} \tbox{A}, \tbox{B}, \tbox{C}). The personalized title of the scientific article being viewed is displayed at the top with the personalized translation underneath it. A slider, labeled from weak to strong personalization, allowed participants to adjust the degree of personalization. The current personalization degree, a numeric value between 0 and 100, is shown just beneath the slider. Participants were instructed to adjust the slider to their preferred degree value before translating using the Translate button. This action triggers an LLM call, which generates and displays a personalized translation.

\subsubsection{History Box}
The history box on the right allows participants to revisit, review, and compare translations they had seen during the session. Each new translation is automatically added to the stack, labeled with its personalization degree and the time of generation. Users can click on any past translation to view it, which would update the translation panel accordingly. They could also mark a translation as a favorite by clicking on the star icon beside it.
Additionally, users can delete translations from the history by clicking the cross button, though we found that participants rarely used this feature in our study.

\subsubsection{Edit Profile}
The edit profile dialog box is accessible by clicking a button on the tool header (\autoref{fig:slider}~\tbox{E}). This displays a short form (\autoref{fig:slider}~[II]) containing information about the participant including their background, age, hobbies, location, and favorite food. This profile information is used to tailor the translation to the user's context. 

\subsubsection{Finishing a Session}
The finish session button (\autoref{fig:slider}~\tbox{F}) served as a proxy measure for their interest in the content after reading the personalized translations. The button triggers a pop-up (\autoref{fig:slider}~[III]) that collects user response on whether they would like to receive the original research paper PDF via email. This button was used at the end of each article's translation session.

\subsection{System Workflow and Implementation}

\begin{figure}[t]
  \includegraphics[width=\textwidth]{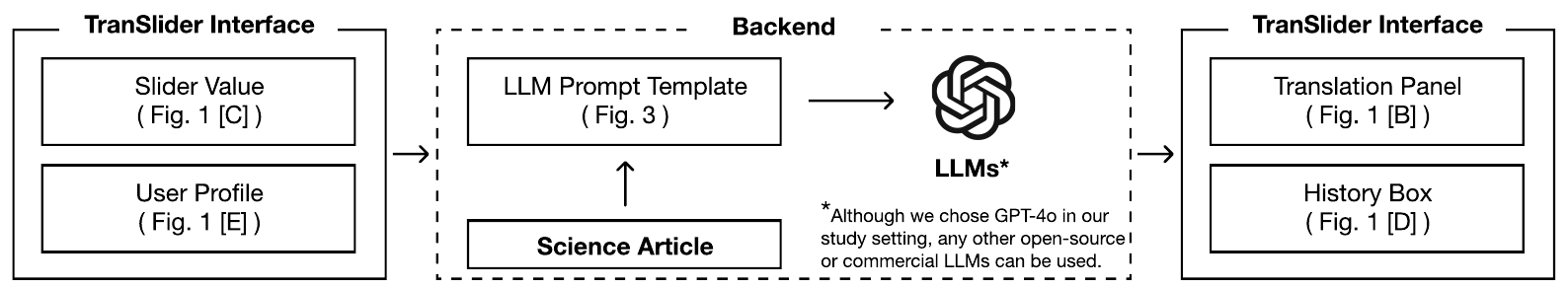}
  \caption{The workflow of \tool{}.}
  \label{fig:workflow}
\end{figure}

\begin{figure}
  \includegraphics[width=\textwidth]{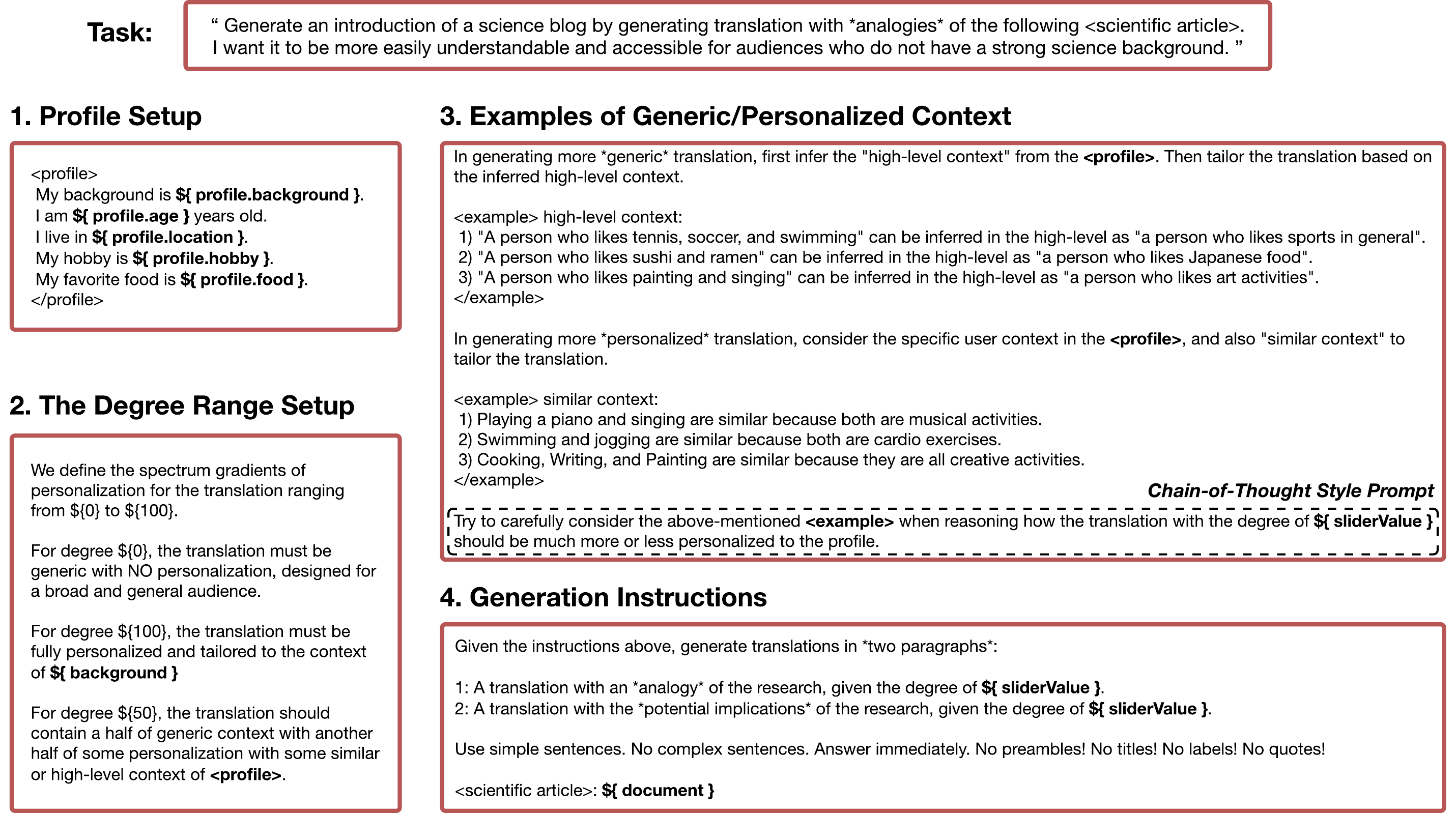}
  \caption{Prompt Template Structure. The prompt begins with a \textbf{Task} overview, instructing the model to use analogies to explain a science article to general audiences. The \textbf{prompt context} consists of four parts: 1. The user profile; 2. Personalization spectrum with descriptions for degrees 0, 50, and 100; 3. In-context examples of both generic and personalized translations; 4. Original scientific article content with the slider value.}
  \label{fig:prompt}
\end{figure}

The workflow of \tool{} is illustrated in~\autoref{fig:workflow}. Once a user adjusts the personalization slider and submits a translation request on the frontend, the backend integrates both the selected personalization degree and the original scientific text into the prompt template (detailed in~\autoref{fig:prompt}). Inspired by existing LLM-powered tools' prompt designs (\eg~\cite{chung2024patchview}), we used the chain-of-thought style instruction~\cite{wei2022chain} to guide the model to reason how the slider input value (personalization degree) should be reflected in the output (personalized translation). With this combined prompt input, the backend makes an API call to the LLM to generate the personalized translation. The API response with the AI-personalized translation is then transmitted back to the translation panel in the frontend. We implemented \tool{} using React.js for the frontend interface and Node.js for the backend. We used OpenAI's GPT-4o model for the LLM. 

\subsubsection*{Note on LLM Model Choice}

While our implementation utilized GPT-4o, this model could be readily replaced with alternative commercial or open-source models. For instance, specialized models like OpenScholar-8B---specifically designed for scientific contexts through retrieval augmentation---outperform GPT-4o for scientific content~\cite{openscholar} and could potentially generate higher-quality personalized translations. Conversely, substituting with older or less capable models such as GPT-Neo~\cite{gpt-neo} would likely result in lower quality and unreliable translations.
Our preliminary testing confirms that alternative models like Meta's Llama-3 (open source)~\cite{llama3}, Anthropic's Claude 3.5 (commercial)~\cite{claude3.5} also generate comparable quality of translations, suggesting that our tool could incorporate different model architectures. Whether there would be any noticeable distinctions in the overall translation quality depending on the model choice is left for future investigation.

\subsection{Pilot Study}

Once the research team was satisfied with the LLM's output, we conducted a pilot study to determine whether the translations were understandable and free from hallucinations or misinformation. We recruited three research scientists outside of our research team. They all had published their work in the past three years. We used their authored publications in Computer Science, Biomedical Engineering, and Electrical Engineering, respectively, for the verification process.

\subsubsection{Procedure} The first author conducted this study in a one-on-one in-person setting. Upon participants' arrival, we introduced \tool{} as a tool designed to help general audiences understand complex scientific texts. We explicitly stated that the purpose of this pilot study was to verify whether the AI-personalized translations were comprehensible and did not hallucinate any misinformation. We then asked participants to provide a link to their recent publication that they would like to share with the general audience. After a brief tutorial of our tool, we had them explore and interact with \tool{} at their own pace without any time constraints. We proceeded to the debriefing interview --- the overall impression of the personalized translations and whether they had seen any misinformation or hallucinations --- only when participants indicated they were ready to share their thoughts. Each of the three pilot studies took about 30 minutes to complete.

\subsubsection{Results} All three scientists found the translations surprisingly understandable. They also confirmed that they could not detect any hallucinations or misinformation in the translations they reviewed. The electrical engineering researcher provided additional feedback on two aspects: the topic selection and the structure of translations. First, they noted that a scientific article's subject matter could influence its perceived benefits to general audiences. Some topics, like health and nutrition, might naturally interest the general audience, while they doubted whether some other topics, like environmental or geological sciences, would draw similar interest. Secondly, they suggested that highlighting some practical implications of studies could make translations more directly beneficial to general audiences.

\subsubsection{Implications} 
\begin{figure*}
  \includegraphics[width=\textwidth]{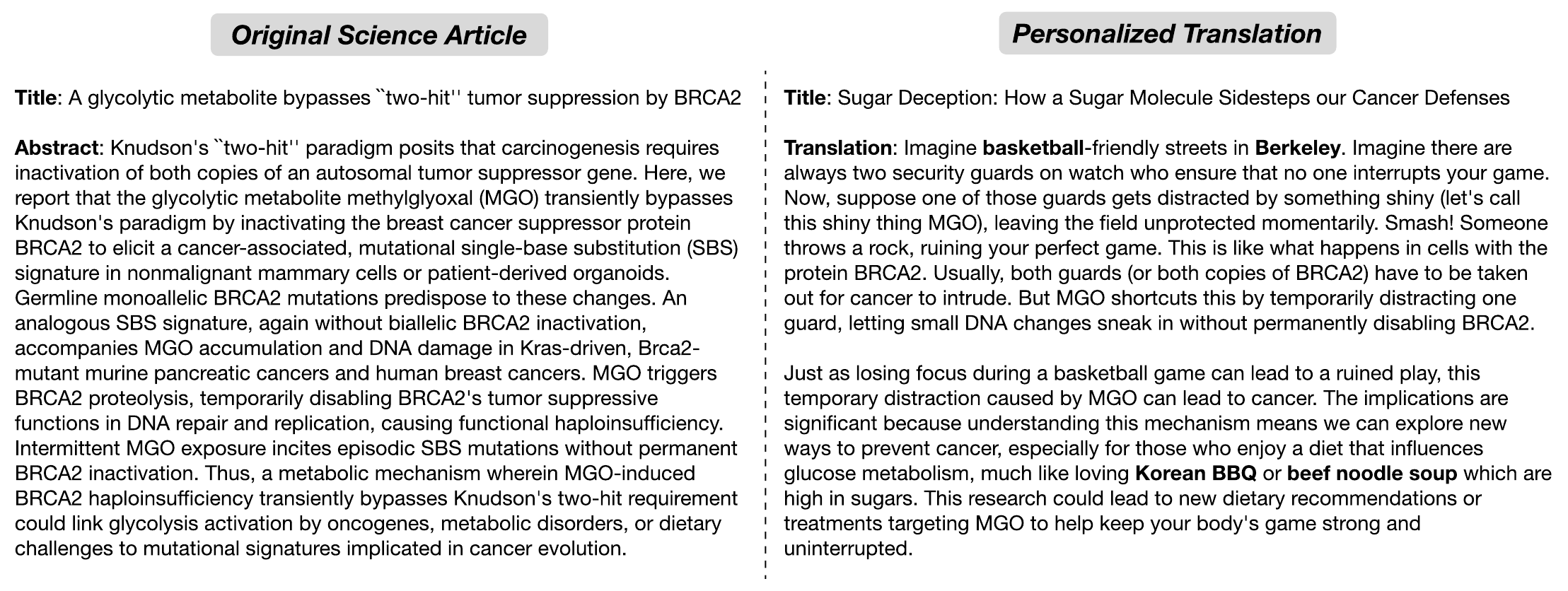}
  \caption{An example translation (degree = 100) of P14, whose hobby is basketball and rock climbing, and lives in San Francisco, CA. Their favorite foods are Korean barbeque and beef noodle soup. The first paragraph gives a brief explanation of the scientific content based on their personal context: location (San Francisco) and hobby (Basketball). The second paragraph presents the implications of this scientific knowledge to the reader's diet and health, saying that both Korean BBQ and beef noodle soup have a lot of sugar.}
  \label{fig:translation}
\end{figure*}

The results of this pilot study influenced our prompt and study design. We refined the instruction prompt (see 4. in~\autoref{fig:prompt}) to generate two distinct paragraphs: the first presenting an analogy explaining the study, and the second discussing implications to the user (example in ~\autoref{fig:translation}). Previously, the translation only explained the study without offering personalized implications. We also decided to include two different types of scientific articles for our study: one with directly relatable content and another with less relatable content (\autoref{study:articlechoice}).

%% file: 4-study.tex
\section{User Study} \label{section:user_study}

Our study aimed to obtain readers' feedback and usage data regarding two aspects of the tool: i) the utility of AI-personalized translation of scientific text (RQ1), and ii) the impact of interactive reading features on user experience (RQ2). Below, we describe the details of our study design, including the choice of science articles, participants, study procedure, and data analysis methods.

\subsection{Science Articles}
\label{study:articlechoice}
We selected scientific articles from two distinct domains: \textit{health} and \textit{environment}. While health-related topics can generate strong interest among broad audiences due to their direct personal relevance (\eg~\cite{jacbos2014cancer, munmun2014seeking, schorch2016designing}), environmental topics might feel less immediately relevant to individuals' personal contexts. We deliberately chose these contrasting domains to demonstrate that the utility of AI-personalized translations may not be dependent on a particular topic; if participants could perceive the value of AI-personalized translations across both more personal (health) and less personal (environmental) topics.

To select science articles that the general audience would be interested in, we searched articles in the Reddit sub-channel r/science\footnote{\url{https://www.reddit.com/r/science/}}. The first author compiled a list of 10 candidate articles for each domain from top-rated posts within the past year. Through iterative discussions among collaborators, we narrowed our selection to two articles: one in health~\cite{kong2024glycolytic} and one in environment~\cite{Guelfo2024}. To mitigate potential ordering effects, we counterbalanced the presentation sequence across participants, with half viewing the health article first followed by the environment article, and vice versa for the remaining participants (last two columns in~\autoref{tab:participants}).

\subsection{Participants}

\begin{table}
\centering
\resizebox{0.95\textwidth}{!}{%
\begin{tabular}{l l l l l l l l l l}
\toprule
    & & & & & \multicolumn{2}{c}{Experience} & \multicolumn{2}{c}{Science Articles}
    \\ \cmidrule(lr){6-7} \cmidrule(lr){8-9}
    
    & & Gender & Age & Education & w/ Sci$^\alpha$ & w/ LLMs$^\beta$ & Session I & Session II \\
    \midrule
    \multirow{7}{*}{A}
      & P1 & Female & 43 & UX Design$^\dag$ & 3 & Yes | No & health & environment \\ 
      & P2 & Male & 23 & Economics & 2 & Yes | No & health & environment \\ 
      & P3 & Male & 32 & Industrial Eng. & 4 & Yes | Yes & health & environment \\ 
      & P4 & Female & 26 & Industrial Eng. & 2 & Yes | No & health & environment \\ 
      & P5 & Female & 24 & Statistics & 2 & Yes | No & health & environment \\ 
      & P6 & Male & 34 & Business  & 1 & Yes | Yes & health & environment \\ 
      & P7 & Female & 24 & Business & 4 & Yes | No & health & environment \\ 
    \hdashline
    \multirow{8}{*}{B}
      & P8 & Male & 25 & Economics & 3 & Yes | No & environment & health \\ 
      & P9 & Female & 25 & Economics & 2 & Yes | No & environment & health \\ 
      & P10 & Female & 23 & Biochemistry & 4 & Yes | Yes & environment & health \\ 
      & P11 & Female & 24 & Computer Sci. & 2 & Yes | No & environment & health \\ 
      & P12 & Female & 23 & Information Sys. & 3 &  Yes | No & environment & health \\ 
      & P13 & Female & 25 & Business & 5 & Yes | No & environment & health \\ 
      & P14 & Male & 24 & Cognitive Sci. & 3 & Yes | Yes & environment & health \\ 
      & P15 & Female & 25 & Business & 2 & Yes | No & environment & health \\ 
\bottomrule
\end{tabular}
}
\caption{Demographic and background of participants. $^\dag$P1 does not have a degree, but has a certificate in UX Design. $^\alpha$Experience reading scientific text in the last three years (1: Never, 2: A few times per year, 3: A few times per month, 4: A few times per week, 5: Daily). $^\beta$Experience with LLMs (first value represents general LLMs experience, second value represents LLM use specifically for reading).}
\label{tab:participants}
\end{table}

We recruited 15 participants (10 female and five male) through mailing lists and word-of-mouth referrals. Our recruitment criteria were designed to represent the general public with an interest in science, but not scientists. We recruited people above 18 years of age and different educational domains---a mix of science-based (\eg engineering) and non-science based fields (\eg business). We excluded people who: (a) had a doctorate degree or a research position, and (b) had a degree in \textit{health} and \textit{environment} fields---the topics of the scientific articles used in our study. This approach allowed us to secure participants with various domain backgrounds with a baseline level of interest in science and unfamiliarity with the chosen articles' fields. We also asked participants how often they read scientific articles (\eg~science blogs) to ensure representation across a spectrum of regular readers and those who rarely engaged with scientific content.

\autoref{tab:participants} illustrates the details of participants' demographics. All participants were fluent in English and aged between 23-43 years (\textit{M=}26.7, \textit{SD=}5.3). Most participants held bachelor's degrees in various fields, 
while one participant (P1) had no degree. In the past three years, eight participants read scientific articles (\eg~science blogs) at least monthly, while the others did so rarely. Regarding LLM experience, all participants were familiar with tools like ChatGPT; four actively used them to understand complex texts (\eg~technical reports or academic papers), while the others did not.

\subsection{Procedure}
\label{study:procedure}
The study was conducted one-on-one and in-person. We audio and screen recorded study sessions that took an hour on average (\textit{M=}60.5 min, \textit{SD=}10.9 min). All participants received a 75 USD virtual gift card as compensation for their time.

\subsubsection{Pre-study Survey (5 min)}
Before conducting the exploratory study, we collected participant background information through a pre-study survey. This included their experience with LLMs and reading scientific texts. We also gathered personal information such as their professional background, hobbies, familiar locations, and favorite foods, which was later used for implicit AI personalization in Session I. 

\subsubsection{Tutorial (15 min)}
After the pre-study survey, we showed the tool interface through a shared monitor in the same room. We explained the tool's purpose and provided a tutorial using a structured script. The tutorial covered tool features such as sliders, buttons, history box items, and included a demonstration of an example interaction. Throughout the tutorial, we encouraged participants to ask questions to ensure their complete understanding of the tool. After the guided tutorial, participants freely explored the tool until they felt comfortable to begin the main study sessions.

\subsubsection{Session I (Implicit) \& II (Explicit) (10 min $\times$ 2)}
\label{sub:session}

As described earlier, all participants conducted the two study sessions in the same order to ensure gradual exposure to the full-featured tool interface. In the first session, participants interacted with the tool version that \textit{implicitly} employs their profile to tailor the scientific articles. We explained to participants that the tool used the profile data they provided in the pre-study survey for personalization. To illustrate this concept, we gave them familiar examples of personalization, such as how Google Chrome leverages users' account data or how social media platforms provide personalized content based on their accounts~\cite{jhaver2023personalizing}. After completing the first session, they began the second session with our fully featured tool that enabled users to \textit{explicitly} edit their profiles. 

In both sessions, we asked participants to freely explore varying degrees of personalization through the slider control. We ensured they reviewed at least five different degrees of personalized translations, and then selected three favorite translations by marking a star to the associated history item. After completing each session, participants responded to a post-session survey, where they rated their familiarity with the scientific articles and wrote the key takeaways about the translated article in 1-2 sentences. We then conducted a post-session interview focusing on: (a) their thought process while interacting with the personalization slider, (b) their rationale for selecting favorite translations, and (c) examples of translations they disliked and their reasons.

\subsubsection{Debrief Interview (20 min)}

After completing both sessions (implicit and explicit), we conducted a semi-structured interview aligned with our research questions. The questions included i) the participants' perceived quality and utility of AI-personalized translations (RQ1), ii) their experience with interactive reading features such as multiple translations, history items, and the slider (RQ2), iii) any particular aspects they liked or disliked, and iv) overall concerns and suggestions for improvement. At the end, participants were free to ask any questions or share additional comments.

\subsection{Analysis}
\label{study:analysis}

\subsubsection{User Behavioral Logs}

We recorded user logs to analyze participants' tool exploration behaviors. The collected data includes user profile information (implicit), profile edits (explicit), explored degrees of personalization, all generated translation texts, favorite translations marked by users, deleted translations, and precise timestamps for all these actions. We analyzed them by plotting and computing descriptive statistics (\eg mean, std, max, and min). We did not conduct a statistical analysis of our data between conditions or groups, because we had a small sample size (\textit{N=}15)~\cite{Gelman2014}. More importantly, our study design was intended to enable qualitative analysis.

\subsubsection{Interview Logs}

We used automated services to transcribe all interview recordings. Then we conducted thematic analysis~\cite{virginia2006using} on the quotes organized by interview question for each participant. For the open-coding process, the first author read quotes and labeled them with initial codes (\ie low-level codes) using Miro board cards\footnote{\url{http://miro.com/}}. Then the first and the second authors collaboratively reviewed the low-level codes to discuss and group codes into potential themes (\ie high-level themes). The first author structured a thematic map showing themes for each corresponding research question. Next, the research team collaborated to refine the thematic map more clearly, build a coherent narrative, and reach a consensus. During this process, we excluded themes that were not directly relevant to our research questions. Ultimately, we finalized the themes for each research question: the utility of AI-personalized translations (RQ1), the impact of interactive reading features on user experience (RQ2), and proposed use cases and associated concerns.

%% file: 5-results.tex
\section{Findings}

We first summarize findings from user behavior log data. We then synthesize participants' interview responses supported by user log data to substantiate additional insights and suggest potential opportunities for future research directions. Our findings cover three key aspects: i) the utility of AI-personalized translation for scientific text, ii) the impact of interactive reading features on user experiences, and iii) the potential use cases and concerns.

\begin{figure}
  \includegraphics[width=0.6\textwidth]{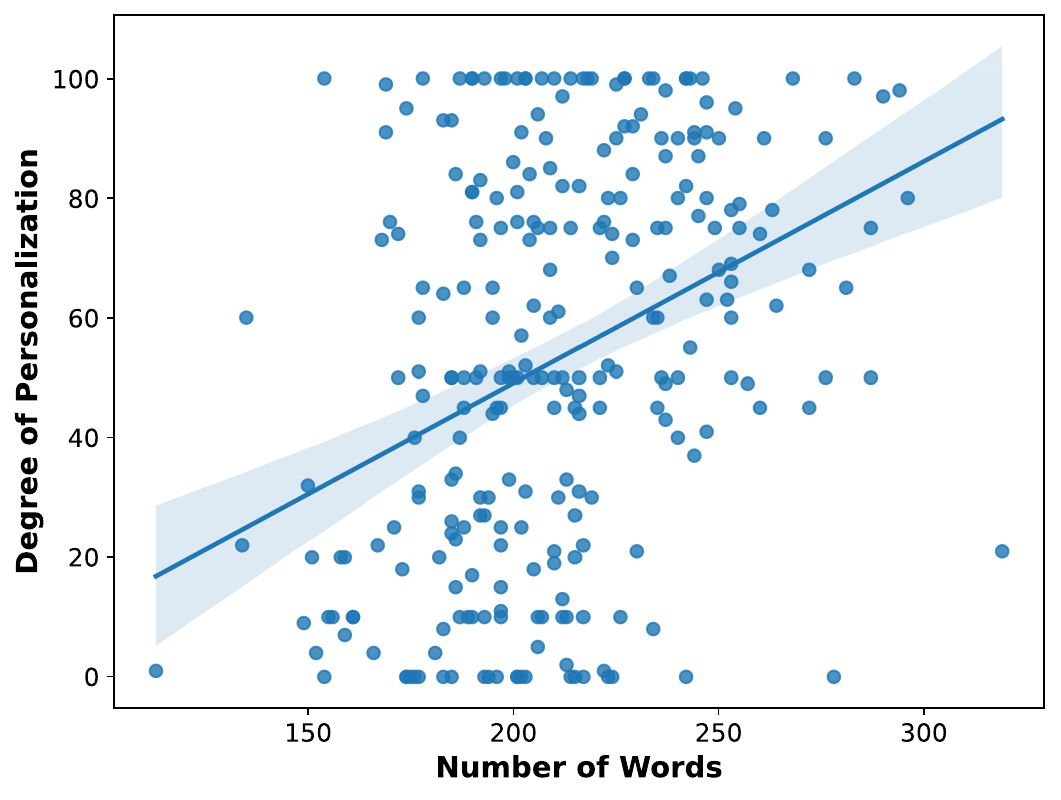}
  \caption{The correlation between the personalization degree and personalized translation length (\#words).}
  \label{fig:correlation}
\end{figure}

\subsection{Quantitative Findings from User Logs and Pre-study Surveys}
\label{finding:quant}

\subsubsection{Users' Exploration Behaviors}
\label{finding:summary}
Participants generated a total of 268 AI-personalized translations using \tool{}, with an average of 8.93 translations explored per session (min: 5, max: 19). We observed a positive correlation (Pearson's $r = 0.36$) between the degree of personalization and length of the translation (\autoref{fig:correlation}), showing that more personalized translations tended to be slightly longer. This subtle increase in length could be attributed to the need for integrating more detailed personal context in highly personalized translations, whereas low personalization required little to no incorporation of personal information.

\begin{figure}
    \centering
    \begin{subfigure}[b]{0.48\textwidth}
        \centering
        \includegraphics[width=\textwidth]{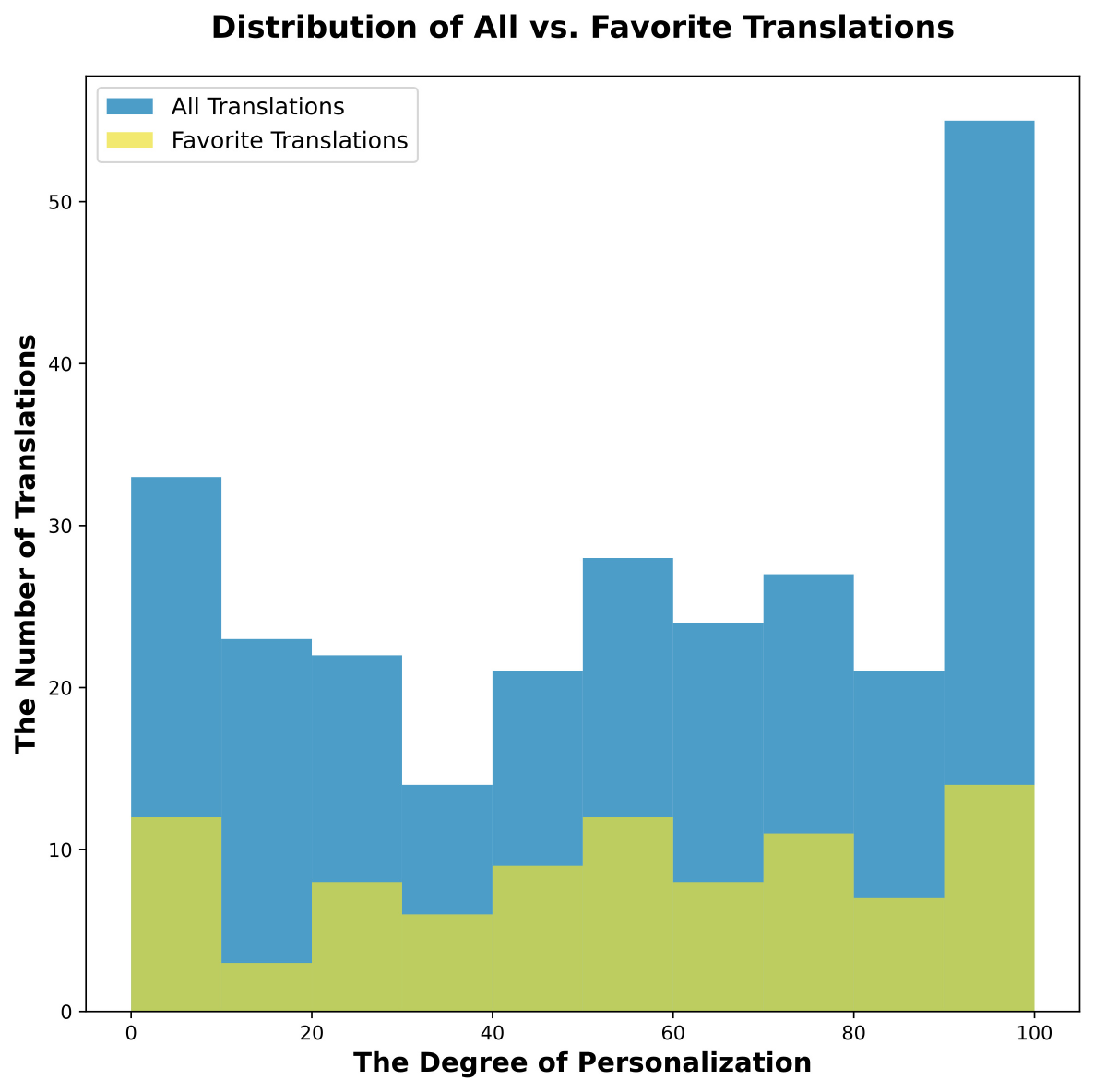}
        \caption{Participants tend to explore extreme personalization degrees (0 or 100) more often, yet their preferred translations typically fell within moderate ranges, despite less exploration of middle ones.}
        \label{fig:favorites}
    \end{subfigure}
    \hfill
    \begin{subfigure}[b]{0.48\textwidth}
        \centering
        \includegraphics[width=\textwidth]{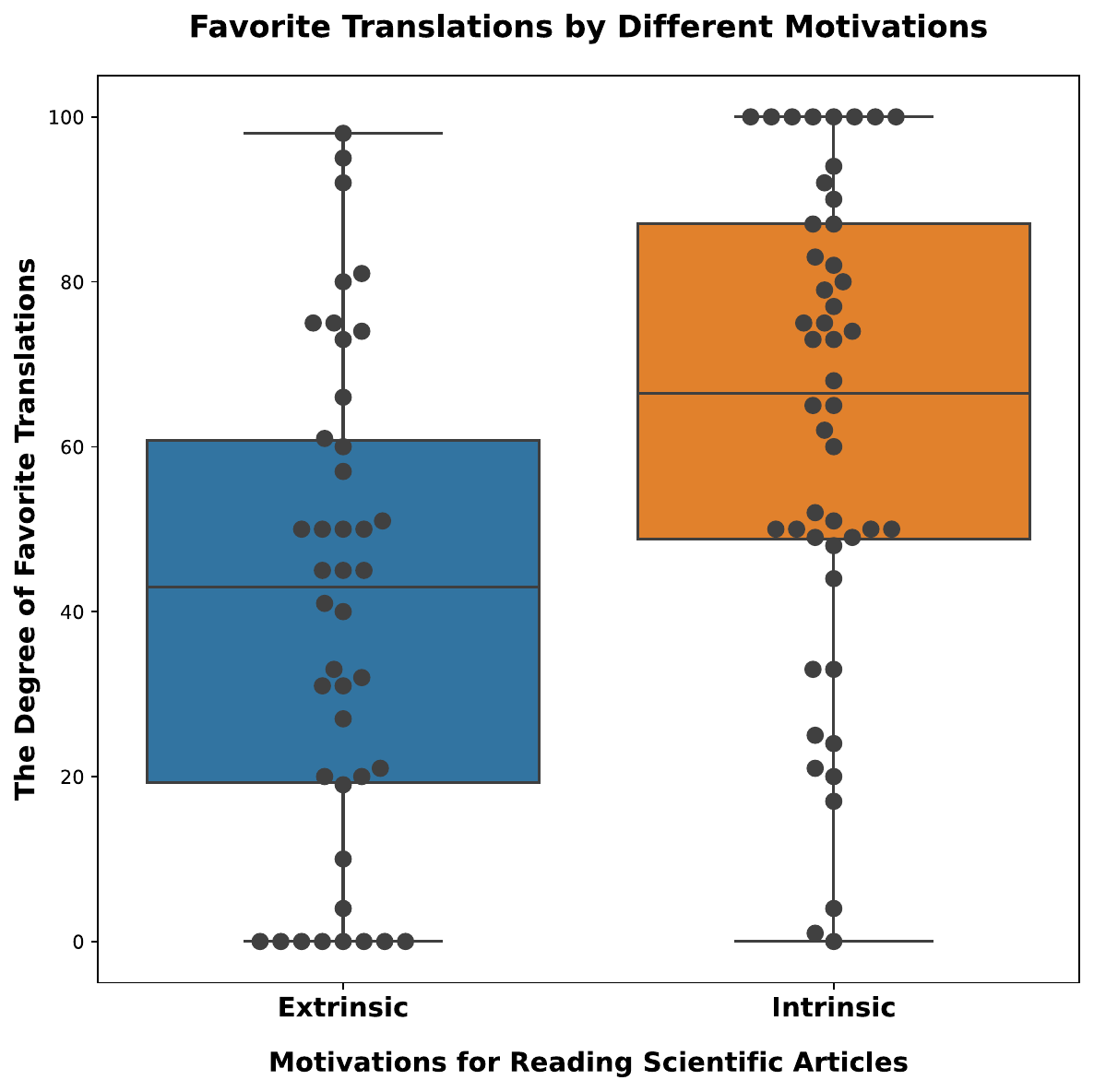}
        \caption{Those with extrinsic motivation for reading scientific articles tend to prefer personalization lower degree, while those with intrinsic motivation like translations with higher degrees.}
        \label{fig:boxplot}
    \end{subfigure}
    \caption{Illustration (a) shows the histogram of two distributions: all translations vs. user-selected translations. Illustration (b) displays a box plot of favorite translations by motivations for reading scientific articles.}
    \label{fig:distributions}
\end{figure}

Participants demonstrated various exploration trajectories, including: i) incremental---starting from low degrees like 2 and moving toward higher degrees, ii) decremental---starting from high degrees like 98 or 100 and moving toward lower degrees, and iii) edge degrees---exploring the extremes (0 or 100) before exploring moderate degrees. Regardless of their starting points, participants often revisited and compared previous translations, moving back and forth between different personalization degrees.
While participants largely explored edge degrees more (0 and 100), many of them eventually favored translations with moderate degrees (\autoref{fig:favorites}). The mean value of participants' favorite translations was 53.14 (\textit{SD=}33.24, \textit{Median=}51.0), indicating an average tendency towards moderate personalization rather than extremes such as hyper-personalization or none at all. However, the high SD shows that participants prefer translations across a wide range of degrees, rather than clustering around a central tendency. This supports our original motivation that singular translations of science articles may not be enough to capture diverse audiences.

\subsubsection{Motivations for Reading Scientific Articles: Extrinsic and Intrinsic}
\label{sum:motivation}

We categorize participants' motivation for scientific articles as extrinsic and intrinsic based on their pre-study interview responses. Those with extrinsic motivations (\textit{N=}7) 
read scientific articles primarily when required for work-related purposes, while intrinsically motivated individuals (\textit{N=}8) 
read them voluntarily, even in their spare time and regardless of work relevance. \autoref{fig:boxplot} illustrates the distribution of participants' preferred translation degrees within each motivation category. Descriptively, we see that participants with extrinsic motivation tend to prefer a lower degree of personalization (\textit{Mean=}40.52, \textit{SD}=29.84), while those with intrinsic motivation favor higher degrees (\textit{Mean=}62.75, \textit{SD=}29.35). While our study was not designed to compare statistical differences between these two groups~\cite{montgomery2018how}, this descriptive result\footnote{We emphasize that our quantitative analysis is limited by the small sample size and participant homogeneity regarding education levels. Hence, it should be treated as preliminary. We encourage a robust analysis before such a system is deployed.} shows that motivation could be an important factor in designing personalization interfaces.

\subsection{Utility of AI-Personalized Translation for Scientific Text (RQ1)}
\label{finding:RQ1}

All participants reported the benefits of personalized translations for scientific articles. However, individuals preferred varying degrees of personalization for different reasons. We asked participants to explain their preferences and rationale with concrete examples they had explored. Based on these examples, we categorized their responses into two groups: those preferring higher degrees of personalization (ranging from 51 to 100) and those preferring lower degrees (ranging from 0 to 45).


\subsubsection{Participants Who Preferred Higher Degree Liked Relatable and Informative Translations}
\label{findings:higher}

\begin{figure}
  \includegraphics[width=\textwidth]{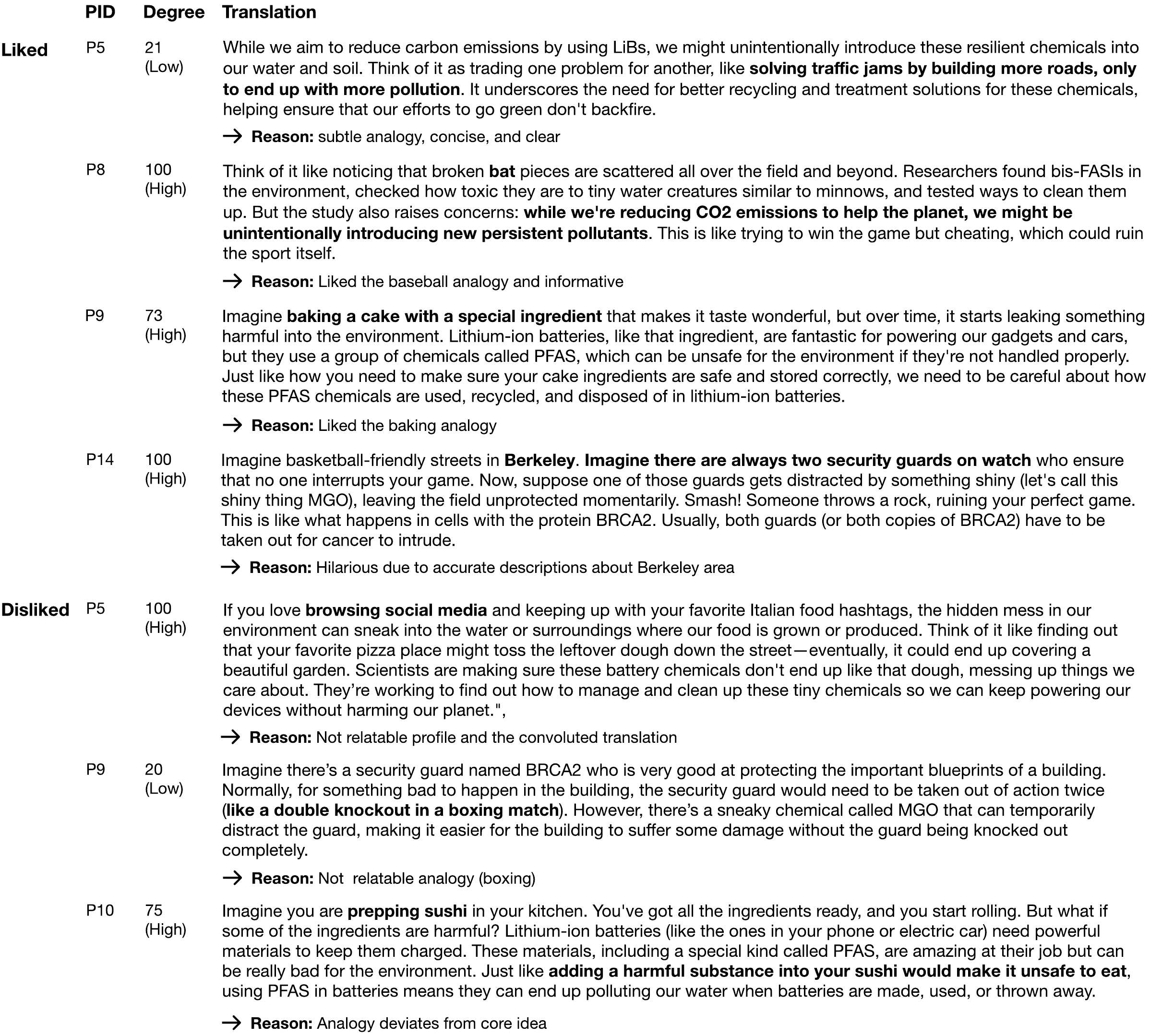}
  \caption{Several examples of preferred translations. While each personalized translation is longer (two paragraphs), we present specific parts that participants pointed out to highlight what they liked or disliked.}
  \label{fig:examples}
\end{figure}

Participants who preferred translations with higher degrees of personalization appreciated that these highly relatable explanations facilitated their understanding of complex ideas (\eg P7, P8, P9, P10, P14). They further stated that these personalized translations, which extensively incorporated participants' personal contexts, resonated with them and were funny. For instance, 
P10 appreciated the translation began with `running', one of their hobbies, making challenging scientific concepts more relatable and easier to understand. Similarly, P14 found some personalizations funny as they incorporated the safety and security aspects of their hometown and cultural background (see translation of P14 in~\autoref{fig:examples}).

\begin{quote}
P10: I like that it started with `running', it brings it up here [in the translation]. But then it continued on the science [content]. Yeah, the science to me is usually the harder part. So I like that. It explained it [science] all at once, and then it started trying to simplify and use analogies. 
\end{quote}

Furthermore, these participants 
felt that highly personalized translations offered detailed scientific content, including additional findings (P8) and detailed explanations of implications (P14).


\begin{quote}
P14: I feel like this last point [implication] was pretty good. Because it says, I'm not sure if the other ones really talked about it that much, but that the cleaner energy could be causing pollution. So the implication of the study is kind of good [useful].
\end{quote}


These participants expressed dissatisfaction with translations of lower degrees. They felt that less personalized translations reduced the depth of scientific information (\eg fewer science concepts and facts addressed in the original article). They also noted that lower degrees of personalization often employ analogies drawn from contexts that were intended to be more universally familiar to general audiences. However, these analogies turned out to be unfamiliar to many participants, ironically making it more challenging to comprehend the scientific articles (\eg P9).


\subsubsection{Participants Who Preferred Lower Degree Liked Concise Translations and Subtle Personalization}

\label{findings:lower}

Some participants (\eg P1, P2, P4, P5) preferred translations with lower degrees of personalization primarily because they demanded less mental load to comprehend. They emphasized the cognitive load required to process long personalized translations, stating that visibly longer translations immediately diminished their interest and willingness to read the material (\eg P2, P5, P13). For instance, translations with lower degrees of personalization tend to be more concise (\autoref{fig:correlation}), which likely contributed to their reduced mental effort. 

\begin{quote}
P5: I feel like I have to put in more effort to understand the analogy in order to understand the actual content. ... it takes a long time, I think for me the length matters. I want it to be personalized, but I don't want it lengthy.
\end{quote}
\begin{quote}
P4: What I liked about less personal ones was that they felt more direct, versus this [high degree personalization] was like reading a lot of text to get to the main point. Yeah, the personalization makes the actual message longer.
\end{quote}

They also appreciated that the analogies were subtly integrated into lower degrees, making them easy to understand and requiring minimal cognitive effort.
\begin{quote}
P5: If I'm ever reading a scientific article, the goal is not to read a lot, but [to figure out] mostly in the shortest amount of time, what is the most information I can take away. So I think having more subtle analogies allowed me to get to the understanding better, but still have the key takeaways.
\end{quote}

\subsubsection{Experimenting with Profile Edits}
\label{rq1:profile}
Despite acknowledging the benefits of personalized translations, regardless of their preferred degrees, participants reported that the translations were sometimes unnatural and difficult to understand. Specifically, they found that certain aspects of their background (personalization profile) were not appropriate to explain the given scientific content. Profile editing enabled them to address this issue by experimenting with different profiles. Participants enjoyed experimenting with their profiles to explore various personalization options. They found it entertaining and fun as they could tell that profile modifications directly influenced the translations. For example, P2 pretended to be a five-year-old Chinese boy, and P3 added ``going to a bar'' as a hobby to receive bar-related analogies.
\begin{quote}
P3: Bar one. I picked the right analogy... And I didn't mean to. I was just like having fun with it. But that's a really good analogy, for PFAS [Per- and Polyfluoroalkyl Substances] I would say.
\end{quote}

Participants noted that hobby-related information influenced the analogies used in translations, while age settings impacted the overall tone.
\begin{quote}
P6: Oh, I thought it was pretty good. I noticed, specifically, when I changed the profile I liked seeing that drastic differentiation. As you mentioned, age was potentially a factor. I'm glad to see that [age] was taken into consideration in the message tone.
\end{quote}

\subsection{Impact of Interactive Reading Features on User Experience (RQ2)}
\label{finding:RQ2}

We analyze participants' responses to two tool features: generating multiple translations per article and slider interaction to control the personalization degree.

\subsubsection{Compounding Understanding with Multiple Translations}
\label{rq2:multiple}

Most participants reported several benefits of reading multiple translations, describing it as assembling a complete picture from partial pieces. They also mentioned that reading multiple translations enabled a cumulative understanding of complex scientific concepts iteratively (\eg P1, P5, P7, P8, P10) and even helped correct initial misunderstandings (\eg P2). For example,


\begin{quote}
P10: As I read more [translations], I understand the topic more. It was helpful as I kept going. Because I feel like I was picking up some small details here and there, and it pieced it together at the end. It definitely wasn't distracting, like completing a big picture like by collecting more positive pieces.
\end{quote}

Furthermore, through an iterative process of generating and reading multiple translations, P2 discovered inconsistencies in their initial understanding. This process helped them identify and correct conceptual misunderstandings that they had not initially recognized, ultimately leading to a more accurate comprehension of the material.
\begin{quote}
P2: I thought the sugar was BRCA2 [BReast CAncer gene 2]. It's actually not. [I noticed from the later translations that] the protein is the BRCA2 and the sugar bypasses the BRCA2.
\end{quote}

However, despite such benefits, a few participants (\eg P2, P3) expressed reluctance to explore multiple translations in realistic scenarios, especially for reading multiple versions of scientific articles that did not align with their interests.

\begin{quote}
P3: In real life, I wouldn't want to read the same thing in a different analogy. If the 1st one didn't work, I'd be like, `Okay, whatever' you know, I wouldn't rerun it. I don't have enough time.
However, I've read different political articles this year, especially with an election year. If there's something I don't understand, I'll go find information elsewhere. In the same way, I would potentially rerun this to see a different analogy or different versions [if I'm interested in the topic]. 
\end{quote}

This shows that articles that deeply interest participants are a strong motivator for them to read multiple translations and gain a better understanding. While \tool{} could offer valuable benefits for comprehension, user engagement may vary depending on individual interests and the perceived relevance of the content.

\subsubsection{Balancing Control vs. Automation Afforded by the Slider}
\label{rq2:slider}

Participants generally recognized and appreciated the gradient differences in personalized translations across the 0-100 spectrum on the slider control. All participants found the slider interaction to be simple, straightforward, and enjoyable, requiring minimal effort to modify the translations of scientific articles.


\begin{quote}
P15: I think that it gave me a lot of control, and I like the control [the agency] that it gave me. I'm selecting what I want to see, when the model is doing [translation] for you. I think that the more you can interact with it [through slider] and kind of control, it makes you feel welcome.
\end{quote}

While participants agreed that the slider was simple and engaging, their desired level of control differed. Some participants expressed interest in having more granular control over the translations (\eg P3, P7, P8, P9, P10), suggesting that the slider alone might not be sufficient. For example, they wanted to manually type specific numbers to modify the degrees of personalization in addition to slider interaction (\eg P9) or prompt interactions to specify more or less personalization.
\begin{quote}
P10: It would be nice if there were an area of more personalization where I tell them what I like about it, and then they do that [personalization] more or do it less. Like, `I would like you [the AI model] to specify less, or I would like you to specify more on either analogy or on the science of it.'
\end{quote}
In particular, P10 proposed a feature to allow users to highlight specific parts of the personalized translations and modify them instead of changing the entire translation (\eg selective editing features~\cite{laban2024beyond}). They envisioned that this feature could enable users to request more detailed elaboration or varying degrees of personalization for each selected part, providing more granular control over how the tool interprets and modifies different text parts.

In contrast, others preferred minimal control or even automated translations without any user intervention. Several participants (\eg P2, P13) found the slider feature overwhelming as it provides too many options (\ie 100 different versions in the range 0 to 100). For example, P13 suggested 10 or even fewer options for personalization rather than a scale of 0-100. Some proposed having just two versions of the article: original and translated, with an on/off toggle.

Furthermore, some participants (\eg P2) suggested a completely automated personalization system. Rather than steering various personalization degrees, they wanted the AI to automatically deliver optimally personalized translations based on user preferences and article topics. Further,   the perceived benefits from the system outweighted privacy concerns.
\begin{quote}
P2: Maybe that [the optimal degree of personalization] should be already part of your profile -- how much personalization do you want. Then it's like, I just want the optimal translation. It's easier for someone to go on a blog and just have it [the optimally personalized translation] straight up there.
\end{quote}

\subsection{Potential Use Cases and Concerns}
\label{finding:concerns}

Participants identified several potential use cases and some concerns. Below, we highlight two use cases: learning unfamiliar content and tailoring their ideas to different audiences. We then present two concerns: the reliability of malleable translations and reduced access to the source article.

\subsubsection{\textsc{Use Case I}: Learning Unfamiliar Content for Various User Motivations}
\label{case:learning}

Participants envisioned using this type of tool to learn unfamiliar topics that they were interested in [intrinsic motivation] such as technology (P1), political pledges (P3), economic content (P10), and how-to documents (P15). Particularly, P10 envisioned using this tool for expanding their knowledge boundaries by learning about unfamiliar or new domains. 

\begin{quote}
P10: I tend to dodge articles about Econ and business. They are not my strong suit. I feel like I would be much more willing to read those [through TranSlider]. Because usually they're filled with so much jargon, so like it now opens up a field for me of something that I'm interested in learning.
\end{quote}

Furthermore, P3 and P9 envisioned using this tool to learn something about an unfamiliar topic (\eg airline compensation policy documents (P3)), which they are not intrinsically motivated to learn, but need to know about [extrinsic motivation].

\begin{quote}
P9: Maybe when I'm trying to find an article that I'm not actually interested in, but I need to get specific information about the topic that I'm not too familiar with where I need the translation. That's where I would see myself using something like this.
\end{quote}

Overall, participants viewed \tool{} as a learning tool to explore various unfamiliar topics and new domains. Moreover, for topics they found less interesting, the tool could help them quickly grasp essential concepts and key takeaways.

\subsubsection{\textsc{Use Case II}: Tailoring Ideas to Communicate to Different Audiences}
\label{case:tailoring}

Participants (\eg P1) also envisioned using this tool to learn how to communicate complex ideas to diverse audiences. For example, the different degrees of personalization could provide tailored translations of how to explain technical products or business ideas to general audiences. P1 envisioned saving multiple translations to their personal library and using them during business meetings.

\begin{quote}
P1: If I need to explain this topic to somebody who's in the industry is very different from how I will explain this to somebody who's not technical. So it's nice to know how I should explain this to somebody who has some knowledge and how to explain it to somebody who has no knowledge like the board of directors for their company.
\end{quote}

\subsubsection{\textsc{Concern I}: Reliability of Malleable Translations from AI}
\label{concern:trust}

While most participants focused more on the benefits, one participant (P8) raised concerns about the reliability of AI-generated personalized translations. They argued that personalized translations can vary with each interaction, raising questions about their consistency and trustworthiness.

\begin{quote}
P8: If I read something and then have to apply that information in my work, it's a little more concerning. I might not trust it because it keeps changing. If I had to cite it, I probably wouldn't trust this. It [AI] might misunderstand something.
\end{quote}

They further stated that this concern would motivate them to check out the original paper for a complete and accurate understanding. While they would not fully rely on the translations in their professional context, they acknowledged that the tool would be useful in deciding whether to read the original document.

\subsubsection{\textsc{Concern II}: Reduced Reference to Original Source Material}
\label{concern:access}

One participant (P5) raised a contrasting concern: she worried that AI-personalized translations might discourage people from referring to original documents. People might rely on these personalized translations rather than engaging with the original material because personalized translations are easier to digest. This could lead to a dependence on AI-processed content at the expense of direct engagement with source materials.

\begin{quote}
P5: The downside is that nobody would probably read the full paper outside of people within that industry or field. But I don't think that's necessarily bad, because the alternative is you have more people understanding key concepts.
\end{quote}


P5 summarized the trade-off: the tool could broaden superficial knowledge access across wider ranging topics but hinder deep understanding and further exploration. Conversely, P5 weighed the risk of shallow learning against the benefit of increased accessibility, noting that without the tool, fewer individuals would gain any exposure to such knowledge.

%% file: 6-discussion.tex
\section{Discussion}

\subsection{Implications for Science Communication and the CSCW Community}

Our study demonstrated the feasibility of addressing the \textit{translation} challenge in science communication. Rather than pursuing a single `optimal' translation, \tool{} generates and enables exploring multiple personalized translations. This approach produced two key benefits: i) a compounding understanding of scientific content, where each translation contributed like a puzzle piece to build a more complete picture of the underlying content, and ii) enabling multi-perspective comprehension, where varied viewpoints conveyed different aspects of the content.

\subsubsection{Towards Collective Understanding of Science with General Audiences}

Our approach could address the \textit{dissemination} issue in science communication: scientists' reluctance to disseminate their work on digital media. Their concerns stem from the potential misinformation arising during translation, which could make audiences misinterpret simplified explanations~\cite{wu2019design, maria2018responsibility}. By enabling readers to explore multiple personalized translations, the \tool{} might create an opportunity for them to build a comprehensive understanding while potentially reducing the risk of misinformation. This approach aligns with established science communication practices, where journalists often craft multiple re-expressions when content might be unfamiliar to general audiences~\cite{hullman2018improving}.

We envision collective knowledge sharing spaces inspired by existing AI-generated content platforms, such as Midjourney~\cite{Midjourney} and DeviantArt~\cite{deviantart}. The Midjourney system shares users' generated images throughout its community by default, allowing individuals to inspire each other to create new and unique creative works along with constructive discussions within the community~\cite{guo2024ex}. Applying similar dynamics to science communication, platforms like Google Scholar~\cite{GoogleScholar} or Semantic Scholar~\cite{SemanticScholar} could integrate our approach to let general audiences explore scientific articles and share personalized translations across their community spaces (\eg~OtherTube~\cite{bhuiyan2022othertube}). Creating such environments 
could foster diverse perspectives and collaborative comprehension of complex scientific materials, and further the larger vision of making science accessible at scale.

Such collective community spaces could also create valuable opportunities for scientists to observe how general audiences digest and interact with their research. Traditionally, scientists have limited visibility into the public perception and potential implications of their work, largely due to a lack of active engagement with broader audiences~\cite{williams2022hci}. However, within this community setting, researchers could preview how individual and collective understandings develop around their work. These insights could influence or engender new research directions. Additionally, such feedback might help scientists identify and even prevent potentially problematic research activities that they would not otherwise anticipate without such collective reflection through public interpretation~\cite{lorente2019scientists}.


However, designers should be aware that malicious actors could exploit such community ecosystems to manipulate the collective understanding of certain topics. Furthermore, some organizations (\eg~`Big Brother'~\cite{orwell20241984}) might leverage these platforms to influence science activities, as public interest often drives funding allocation, and without proper financial support, some research endeavors may not remain sustainable. We see this line of research as a promising new direction.

\subsubsection{Enhancing Communication by Surfacing Diverse Perspectives}

The benefit of having varied perspectives could be useful in contexts such as education and cross-disciplinary collaboration. Education has grown increasingly complex, particularly with the widespread adoption of online learning~\cite{li2021challenges}. While online platforms provide greater scalability, instructors face persistent challenges in delivering effective instruction to large, diverse student cohorts~\cite{kulkarni2015}. AI-powered personalization tools could help address these challenges by offering instructors tailored translation suggestions adapted to different students.
For example, systems could let children input their favorite animals, foods, or objects relevant to the given learning context, and generate tailored explanations and learning materials to better engage students. Future research should explore the impact of these personalized translations on learning outcomes, such as learning performance.

Likewise, in cross-disciplinary contexts, effective communication is often challenging due to several obstacles, such as divergent terminology~\cite{hou2017hacking}, varying data interpretations~\cite{kim2022putting}, and different motivations~\cite{mao2019data}. Research shows that new ideas are better understood when presented through familiar examples~\cite{kim2022putting}. Cross-disciplinary collaborative activities (\eg drafting documents, building systems, and shaping policy letters) require clear and constant communication among collaborators~\cite{saha2021urban, yoo2024missed}. While real-time conversations offer opportunities to address misunderstandings immediately, scheduling synchronous meetings among collaborators inevitably creates communication delays that can slow project progress. Our interface could benefit such asynchronous communication scenarios by allowing collaborators to generate personalized translations of complex content, helping team members from different disciplinary backgrounds understand shared documents without waiting for clarification meetings.

However, these approaches may lead to over-reliance on AI-personalization, as indicated by a participant (P5). In a future where AI tools work exceptionally well at addressing communication barriers, individuals might prefer interacting with these interfaces rather than directly communicating with students, instructors, or collaborators. The convenience of AI-generated `perfect' translations might even diminish individuals' efforts and motivation to develop clear communication skills and articulate their ideas in accessible ways. This raises crucial concerns about balancing between technological assistance and the development of fundamental human agency in communication capabilities, suggesting that such tools should be designed to augment rather than replace human roles. Further research is warranted on how to design and develop such AI-driven cross-disciplinary collaborative interfaces while balancing the perceived benefits, potential long-term harms, and ethical concerns.

\subsection{Interface Design Enabling User Control for Human-AI Alignment}

We designed an interactive interface that empowers users to steer and align AI outputs to their preferences, as opposed to optimizing the AI models directly~\cite{ouyang2022follow}. While many participants appreciated the controllability of AI output, we found some users may not want such control in-the-wild. Participants also showed varied responses about the reliability of AI outputs.

\subsubsection{Flexible AI Interfaces for Balancing User Control and Automation}

Our interface provides three key features to support user agency: a) a slider that allows users to adjust the degree of personalization, b) a profile editing function, and c) a history box. These features enable users to control AI outputs without technical expertise, save personalized preferences, and iteratively explore various outputs until they find desirable results. While highly accurate AI models might better serve diverse user needs, achieving such accuracy often requires extensive user data collection, raising concerns of cost, privacy, and selection bias amongst annotators~\cite{sambasivan2021everyone}. Our design approach provides an alternative: rather than relying on AI-powered systems to perfectly model the plurality of user preferences, we empower users to actively align AI outputs to their individual preferences. This approach not only preserves user privacy but also maintains human agency in human-AI interaction, addressing critical concerns about AI overreliance in our society~\cite{zana2021trust, Okolo2024}.

However, we should acknowledge a potential limitation of our approach. Our interaction design assumes users want to preserve their agency over AI systems, which may not always be true depending on individuals and contexts. Indeed, a participant (P2) stated they do not want to interact with systems in-the-wild (\autoref{rq2:slider}). As our study suggests, in designing systems for mass audiences, we must recognize that a portion of users, like P2, may prefer fully automatic personalization without manual adjustments. Also, in different contexts such as entertainment and creative art domains, users may prefer consuming the contents passively, rather than proactively interacting with systems~\cite{kim2024authors}. Therefore, future systems should balance automation and user control, perhaps by offering flexible interfaces (\eg~\cite{min2025malleable}) providing both \textit{automatic} and \textit{control} modes that users can switch between depending on their needs.

\subsubsection{The Reliability of AI-Personalized Translations}

Participants expressed contrasting concerns about the reliability of AI-generated translations for science communication. While some participants were reluctant to rely on the translations, others were worried that people might depend too heavily on them. As AI technology becomes more sophisticated, the higher accuracy of translations with fewer instances of incorrect or hallucinated content might mitigate such reliability concerns~\cite{wu2017automantic}. However, regardless of translation quality, concerns about AI-processed output and the trustworthiness of AI may persist. The fact that AI-generated translations are not human-authored continues to affect the perceived trust~\cite{Ryan2020}. The perceived mistrust could stem from an ambiguity in responsibility, specifically regarding the onus for mistranslations and their resultant ramifications. For instance, a participant mentioned that they would not trust AI outputs for work-related tasks without verifying the source, as they are personally accountable for any errors in the work.

To mitigate risks, verifying translations against original source materials could be an approach. However, who should bear the responsibility: readers or authors? Putting the onus on the reader defeats the purpose of personalization for making science accessible as original scientific articles use overly complex language and jargon. Our findings and prior work indicate that readers do not want to invest significant mental effort parsing complex texts, resulting in the risk of over reliance on AI-generated translations~\cite{zana2021trust}. A potential solution could be for authors (\ie~scientists) to verify several versions of AI translations before publishing to the general public~\cite{kim2024authors}. For example, scientists could go through 5--10 different translations with varying degrees of personalization and approve a subset as verified translations. The balance between utilizing AI translations and encouraging verification of original sources presents a unique challenge, particularly given the current state of scientific communication. Further explorations are needed to find scalable approaches such that the author verification does not become cumbersome.

\subsection{Limitations and Future Work}

Our study had several limitations. First, it was conducted in a small controlled setting. Specifically, a pool of 15 participants were asked to explore two pre-selected science articles from health and environment domains using our tool. Future studies should consider a large-scale in-the-wild study where participants could choose from a wider range of scientific articles. Secondly, our participants might not fully represent the general audience with varying degrees of formal education. More research is needed by studying participants with diverse educational levels (\eg~high school graduates or those with lower educational attainment), to better understand how people with lesser exposure to formal education could draw benefit from our tool. Similarly, our participant demographics were limited, with most participants (12/15) in their twenties and recent college graduates. Only three participants were from other age groups. This age-homogeneous participant pool may have influenced the findings of our study. Future studies should include participants from diverse backgrounds in terms of age, education level, and culture to capture a broader range of perspectives on AI-personalized translation tools. For instance, the impact of AI-personalized translation tools in high school science learning classes~\cite{Lee_Perret_2022} or children-parent learning contexts~\cite{cingel2017how} could be different from our findings. Lastly, our study setup required participants to provide feedback based on brief interactions with the tool. The long-term effects of interactive reading experiences with AI-personalized translation tools may differ from our findings. Future longitudinal studies are warranted to provide a more comprehensive understanding of how users interact with and benefit from these types of tools over extended periods. 

%% file: 7-conclusion.tex
\section{Conclusion}

Our study investigated the potential of AI-personalized translations to enhance science communication in digital media platforms. We developed \tool{}, an AI-powered reading interface that generates personalized translations of scientific text for general audiences. The tool's key feature involves an interactive slider that allows users to adjust the degree of personalization to generate tailored translations. Through an exploratory study with 15 participants, we found that participants preferred varying degrees of personalization for the tailored translations, from deeply relatable and contextualized translations to short and slightly contextualized ones. Interestingly, reading multiple translations enabled them to have a compounding understanding of scientific content. We present several implications for science communication and for supporting communication across individuals with diverse backgrounds in collaborative contexts. We also discuss designing flexible interfaces that allow users to preserve control over AI tools to steer desired outcomes. We hope our study opens a new avenue to promote science communication for general audiences.